\documentclass[12pt]{article}
\usepackage{graphicx,amssymb,amsmath,amsthm,empheq,color}
\usepackage[colorlinks=true,urlcolor=blue,linkcolor=black,citecolor=black]{hyperref}
\usepackage[nointegrals]{wasysym}
\usepackage{authblk}

\makeatletter

\newlength{\fighskip} \fighskip=2pt
\newlength{\figvskip} \figvskip=3pt

\newcommand*{\wideboxed}[1]{\setlength{\fboxsep}{1ex}%
  \fbox{\m@th$\displaystyle#1$}}

\def\ubrace#1_#2{%
  \underbrace{#1}_{\hb@xt@\z@{\hss$\scriptstyle#2$\hss}}}

\newcommand{\ph}{\varphi}
\newcommand{\kap}{\varkappa}

\newcommand{\RR}{\mathbb{R}}

\DeclareMathOperator{\erf}{erf}
\DeclareMathOperator*{\EE}{\mathbb{E}}
\newcommand{\const}{\mathrm{const}}

\newcommand{\init}{\mathrm{in}}

\newcommand{\abar}{\bar{a}}
\newcommand{\Kelly}{\mathrm{Kelly}}

\newtheorem{Proposition}{Proposition}

\makeatother

\title{Calculated Boldness \\
 \large Optimizing Financial Decisions with Illiquid Assets}
\author[1]{Stanislav Shalunov}
\author[2]{Alexei Kitaev}
\author[2]{Yakov Shalunov}
\author[3]{Arseniy Akopyan}

\affil[1]{\normalsize\it FORA Capital}
\affil[2]{\normalsize\it California Institute of Technology, Pasadena, CA 91125, USA}
\affil[3]{\normalsize\it IITP RAS (Kharkevich Institute), Moscow, Russia}

\date{December 23, 2020}

\begin{document}
\maketitle

\begin{abstract}
We consider games of chance played by someone with external capital that cannot be applied to the game and determine how this affects risk-adjusted optimal betting. Specifically, we focus on Kelly optimization as a metric, optimizing the expected logarithm of total capital including both capital in play and the external capital. For games with multiple rounds, we determine the optimal strategy through dynamic programming and construct a close approximation through the WKB method. The strategy can be described in terms of short-term utility functions, with risk aversion depending on the ratio of the amount in the game to the external money. Thus, a rational player's behavior varies between conservative play that approaches Kelly strategy as they are able to invest a larger fraction of total wealth and extremely aggressive play that maximizes linear expectation when a larger portion of their capital is locked away. Because you always have expected future productivity to account for as external resources, this goes counter to the conventional wisdom that super-Kelly betting is a ruinous proposition.
\end{abstract}

\section{Introduction}

Suppose you---the reader---have arrived at a conference in Los Vegas and are offered an opportunity. You are given $\$1{,}000$ and the chance to play a special game. The game is simple: you can stake as much of the money on a coin flip as you want, but no other money you may have. If you win, you earn $+0.3$ of whatever you put down. If you lose, it's $-0.25$ of the stake. Then you do it $999$ more times.

If you wanted to, you could just walk away with the $\$1{,}000$ and go about your day. But this is a positive expectation game. On average, you win $+0.025x$ every game. With $1{,}000$ games, the expected returns of putting down all your available money on every flip would be approximately $\$53{,}000{,}000{,}000{,}000$. Astronomical, but practically imaginary: if you bet that way, half the time you'll get nothing, less than a measly $\cent\kern1pt 0.3$. If you wish to walk away with more than your starting $\$1{,}000$, your odds are a mere $7\%$ (with a fixed fraction bet, the order of outcomes is irrelevant; thus, if $k$ is the number of winning flips, $k < 524$ gives less than $\$1{,}000$ at the end, and a simple binomial distribution lets you calculate the probability). All your winnings are concentrated in a small number of unlikely, massive outcomes. Not playing at all is wasteful, but betting everything is too risky. This raises an obvious question: how \textit{should} you play?

This is not a meaningless hypothetical: we are not considering this game arbitrarily but rather because it is a simplified model of life. We can view career progression as a series of decisions about how much effort to invest in opportunities as they come up. Furthermore, since overall wealth is growing, we can assume that these opportunities average to a positive expectation. However, since just investing everything blindly in each one obviously leads to ruin, we know the median result of opportunities is negative. So we come to the rough conclusion that life is a series of rounds, each with a positive expectation but negative median result, making this game a reasonable proxy and an interesting problem to solve.

To account for the cost of risk and variance, the best option is not to optimize expected returns but to optimize the value of some utility function. We go into more detail on utility functions in the next section, but the one which we will be studying is the logarithm. It was first proposed in~1738 by Bernoulli~\cite{Bernoulli}, and its modern use in repeated gambling and investment is credited to Kelly~\cite{Kelly56}. If you simply optimize the logarithm of returns, you get the far more reasonable strategy of placing $\frac{1}{3}$ of your capital on every flip. Your probability of walking away with more than $\$1{,}000$ skyrockets to $93\%$ and the median value increases to roughly $\$63{,}000$.

However, this strategy is still suboptimal. After all, if Kelly optimization gives you the optimal risk profile for a given wealth, you have to account for total wealth, including wealth other than the $\$1{,}000$ you've been given to play with. Even if you are completely broke, you have all your future earnings and prospects to factor in to play optimally. In other words, you need to optimize $\ln(1+x)$ (normalizing external capital to $1$), a problem far more complicated than simply optimizing $\ln(x)$ (simply betting $\frac{1}{3}$) and the one we will be solving in this paper.

Part of conventional wisdom is that reasonable behavior is at least as risk-averse as a logarithm and that less risk aversion leads to ruin. Using the family of isoelastic utility functions (see section~\ref{sec:isoelastic}) we can see this intuitively by noting that logarithm and all more risk-averse functions give a value of $-\infty$ at $0$ while anything less risk-averse than logarithm gives it a value of $0$. It's easy to see that no bet that leads to a possibility of ruin will ever be taken in the former case, while in the latter case bets with a chance of ruin can be taken, meaning the probability of bankruptcy approaches $1$ with time.

However, one of the key results we see here is that cases where it is rational to bet more than the ``Kelly bet'' not only exist in life but are actually perfectly normal, even for individuals whose overall utility function places true ruin at $-\infty$.

\section{Utility functions}

We have already mentioned Bernoulli's thesis: for a person with total wealth $w$, the utility of this sum is given by $\ln w$. In any gamble, a rational player should maximize the expectation value of this quantity (but not, for example, the expectation value of $w$ or $\sqrt{w}$). However, it appears that such a sweeping statement cannot be true because different people have different preferences about acceptable risks and how much money they need. For example, $-\frac{1}{w}$ appears to represent behavior of some risk-averse individuals well. An objective analysis of different utility functions is achieved by focusing on the corresponding optimal behaviors and their outcomes, which can be compared using some universal measures. 

\subsection{Isoelastic Utility Functions}\label{sec:isoelastic}

In a deterministic situation, the utility can be defined by any monotonically increasing function $u(w)$, and maximizing it is equivalent to maximizing $w$. When chance is involved, it is natural to consider concave functions because attaining a certain wealth $w$ is usually better than expecting a random gain with the same average. (In any case, the first option can be converted to the second by going to a casino.) Among monotone, concave functions, we focus on functions of the form $u(x)=cx^{\alpha}+\const$. The constant term and the coefficient $c$ do not matter as long as $c$ has the correct sign, that is, positive if $0<\alpha<1$ and negative if $\alpha<0$. Let us also include the $\alpha=0$ case:
\begin{equation}
u_{\alpha}(x)=\begin{dcases}
\alpha^{-1}(x^{\alpha}-1)
& \text{if } \alpha\in(-\infty,0)\cup(0,1],\\
\ln x = \lim_{s\to 0}s^{-1}(x^{s}-1)
& \text{if } \alpha=0.
\end{dcases}
\end{equation}
This is termed the ``isoelastic utility function with risk aversion $1-\alpha$''. We prefer to work with $\alpha$ and call it the \emph{risk parameter}. It will be shown that using $\alpha>0$ can be very risky but $\alpha\le 0$ is relatively safe.

Isoelastic utility functions lead to myopic optimal strategies (strategies that do not consider history or future opportunities and treat each opportunity independently), which makes them practically applicable.

\begin{figure}
\centerline{\includegraphics{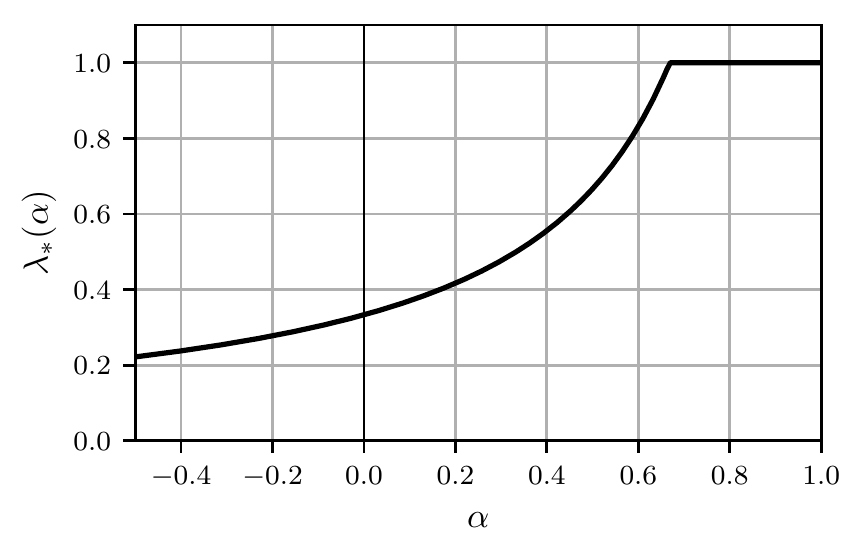}}
\caption{Optimal betting fraction $\lambda_*$ as a function of the risk parameter $\alpha$ for the gamble with gain factors $a_1=1.3$,\, $a_2=0.75$ and probabilities $p_1=p_2=0.5$. In this example, the Kelly fraction is $\lambda_*(0)=1/3$, and the gamble is attractive (i.e.\ $\lambda_*(\alpha)=1$) for $\alpha\ge\alpha_1\approx 0.6685$.}
\label{fig:opt_frac}
\end{figure} 

Let us discuss the maximization of $\EE[u_\alpha(w)]$ in a one-time gamble. The player's decision is represented by a single parameter $\lambda\in[0,1]$, the fraction of his initial wealth $w_\init$ he is willing to bet. The game is defined by some gain factors $a_j$ and probabilities $p_j$. (For example, each round of the game described in the introduction has $a_1=1.3$,\, $a_2=0.75$, and $p_1=p_2=0.5$.) Thus, the player will own the amount
\begin{equation}
w_j(\lambda)=(1+\lambda(a_j-1))w_\init
\end{equation}
with probability $p_j$. Adopting the utility function $u_\alpha$, we are interested in maximizing its expectation value,
\begin{equation}
U_\alpha(\lambda) = \EE_{j}\bigl[u_\alpha(w_j(\lambda))\bigr]
=\sum_j p_j u_\alpha(w_j(\lambda)).
\end{equation}
This is a concave function of $\lambda$, and its derivative is monotonically decreasing:
\begin{equation}
U_\alpha'(\lambda) = w_\init^{\alpha}\sum_j p_j
\frac{a_j-1}{(1+\lambda(a_j-1))^{1-\alpha}}.
\end{equation}
Therefore, there are three possibilities for the value of $\lambda=\lambda_*$ at which $U_\alpha(\lambda)$ attains its maximum:
\begin{equation}
\begin{array}{cl@{\qquad}l}
\lambda_*=0 & \text{if }\, U_\alpha'(0)=w_\init^{\alpha}\sum_{j}p_j(a_j-1)\le 0
& \text{(unfavorable gamble);}\vspace{3pt}\\
\lambda_*=1 & \text{if }\,
U_\alpha'(1)=w_\init^{\alpha}\sum_j p_j(a_j^\alpha-a_j^{\alpha-1})\ge 0
& \text{(attractive gamble);}\vspace{3pt}\\
0<\lambda_*<1 & \text{otherwise,\, in which case } U_\alpha'(\lambda_*)=0
& \text{(intermediate case).}
\end{array}
\end{equation}
Note that ``favorability'' is a property of the gamble itself; it simply means that the average gain factor $\abar=\sum_jp_ja_j$ is greater than $1$. ``Attractiveness'', on the other hand, depends on the risk parameter $\alpha$. See figure~\ref{fig:opt_frac} for illustration.

Now, we will examine the risks the utility function $u_\alpha$ entails, particularly, for $\alpha>0$. It is sufficient to consider only attractive gambles because a $\lambda_*$ bet is equivalent to betting all of one's money in a modified game, where $\tilde{a}_j=1+\lambda_*(a_j-1)$. For games with two possible outcomes, we have the following constraints:
\begin{equation}
p_1(a_1^{\alpha}-a_1^{\alpha-1})+p_2(a_2^{\alpha}-a_2^{\alpha-1})\ge 0,\qquad
p_1+p_2=1.
\end{equation}
They can be satisfied for arbitrarily small positive numbers $p_1$, $a_2$ by choosing a sufficiently large $a_1$. (Here we have assumed that $\alpha>0$.) Thus, under certain circumstances, the player will risk an arbitrarily large loss with probability arbitrarily close to $1$. For example, a player with utility function $u_\alpha$,\, $\alpha=1/2$ will fall for this proposition: $a_1=1000$,\, $a_2=0.1$,\, $p_1=0.1$,\, $p_2=0.9$, which implies a tenfold loss with $0.9$ probability.

Bernoulli's utility function ($\alpha=0$) helps avoid such reckless behavior. In this case, the attractiveness condition becomes $\sum_{j}p_ja_j^{-1}\le 1$, or simply $p_2\le a_2$ for two-outcome games with $a_1\to\infty$. Thus, a Bernoulli player will risk a loss by factor of $A$ if it occurs with probability $1/A$ or less. That is what we called ``relatively safe''. However, the risk increases if fractional bets are not allowed. Here we come to an implicit assumption on which Bernoulli's theory rests---the possibility to manage risks by splitting capital (e.g.\ into a bet and a safe portion). This concept is formalized below as a convex game.

\subsection{Log Optimization}

A general game of chance is described by a collection of positive real numbers $w_J(\Lambda)$, where $\Lambda$ represents a playing strategy and $J$ some random event, occurring with probability $p_J$. This definition is suitable as a description of each gambling round as well as the game as a whole. In the latter case, $\Lambda$ is some function that prescribes player's actions throughout the game. We assume that the randomness is external to the player, who makes deterministic decisions based on available information. We are interested in optimizing the average Bernoulli utility,
\begin{equation}
U_0(\Lambda)=\EE_J\bigl[\ln(w_J(\Lambda))\bigr] =\sum_J p_J \ln(w_J(\Lambda)).
\end{equation}
The strategy $\Lambda$ maximizing $U_0(\Lambda)$ is called the Kelly strategy, though in his paper, Kelly focused on repeated gambling. 

A \emph{convex game} enjoys the property that for any strategies $\Lambda^{(0)},\Lambda^{(1)}$ and any number $0\le t\le 1$, there is an \emph{interpolating} strategy $\Lambda^{(t)}$, namely, one satisfying the condition
\begin{equation}
w_J\bigl(\Lambda^{(t)}\bigr)
=(1-t)\,w_J\bigl(\Lambda^{(0)}\bigr)+t\,w_J\bigl(\Lambda^{(1)}\bigr)\quad
\text{for all } J.
\end{equation}
Logarithmic optimization for convex games, and, in particular, games based on investment portfolios, was studied in~\cite{BeCo80,BeCo88}. These papers, among several others, demonstrate that the Kelly strategy is superior not only in terms of maximizing a particular function, but in a more objective sense. We now discuss some simple properties of this kind. The following result (in the context of portfolios) appears as Corollary~2 in Ref.~\cite{BeCo80}.

\begin{Proposition}\label{prop:compstrat}
Let $\Lambda_*$ and $\Lambda$ be, respectively, the Kelly-optimal strategy and an arbitrary strategy in a convex game, and let $A\ge 1$. Then the probability of the event $w_J(\Lambda)/w_J(\Lambda_*)\ge A$ is at most $1/A$.
\end{Proposition}

\begin{proof}
Consider the strategy $\Lambda^{(t)}$ interpolating between $\Lambda_*$ and $\Lambda$. Since $\Lambda_*$ is optimal, we have
\begin{equation}
0\ge \frac{dU_0\bigl(\Lambda^{(t)}\bigr)}{dt}\bigg|_{t=0}
=\sum_{J}p_J\frac{w_J(\Lambda)-w_J(\Lambda_*)}{w_J(\Lambda_*)}
=\EE_{J}\biggl[\frac{w_J(\Lambda)}{w_J(\Lambda_*)}\biggr]-1.
\end{equation}
The statement in question follows immediately by the Markov inequality.
\end{proof}

Essentially, strategy $\Lambda_*$ is hard to beat by a large factor. This can be compared intuitively to the idea that if you have some amount of money, say $\$1$, and any form of non-advantageous casino, you will be able to obtain $\$A$ with probability no greater than $1/A$. Obviously, playing in a non-advantageous casino is hardly a good idea.

A key observation in Kelly's original paper~\cite{Kelly56} is that for repeated gambling or investments, maximizing the logarithmic utility is the same as maximizing growth rate. This can be shown rather simply. If $w_n$ is the wealth after round $n$ (for $n=0,1,2,\ldots$), then the growth rate is defined as $\lim_{n\to\infty}\frac{1}{n}\ln\frac{w_{n}}{w_{0}}$. For any myopic strategy, this limit is equal to $\EE\bigl[\ln\bigl(\frac{w_{n+1}}{w_n}\bigr)\bigr] =\EE[\ln w_{n+1}] - \EE[\ln w_n]$ with probability $1$. This follows from the strong law of large numbers. Note that maximizing $\EE\bigl[\ln\bigl(\frac{w_{n+1}}{w_n}\bigr)\bigr]$ (and thus, the growth rate) is different from maximizing the expectation value of $\frac{w_{n+1}}{w_n}$.

The Kelly strategy dominates any other strategy in the long run. A naive statement of this property would be that if $w^*_n$ and $w_n$ are the results of playing by the Kelly strategy and any other strategy, respectively, then $\Pr[w^*_n\ge w_n] \to 1$ as $n\to\infty$. But this is true only for myopic strategies; in the general case, Thorp gives a counterexample~\cite{Thorp11}. There are different ways to formulate asymptotic optimality for general strategies, see e.g.~\cite{AlCo88}.

Because of these advantages, we consider specifically Kelly optimization with external capital, rather than optimization of some other utility function.

\section{Repeated gambling and dynamic programming}

Now we define the main problem formally. The game consists of $n$ identical rounds. We count them from the end, since the important factors in making a decision is the remaining number of rounds $k$ and the current amount $x_k$. (Our notation is a bit complex as we keep record of the game history.) The initial capital $x_n=x$ is fixed. Each round is a favorable gamble with gain factors $a_j$ and probabilities $p_j$. The player can bet any fraction $\lambda_k\in[0,1]$ of the current amount. Depending on the chance event $j=j_k$ occurring with probability $p_{j_k}$, the new amount is
\begin{equation}\label{x_update}
x_{k-1}=(1+\lambda_k(a_{j_k}-1))x_k.
\end{equation}
A playing strategy is a sequence of functions $\Lambda=(\Lambda_n,\dots,\Lambda_1)$, where $\Lambda_k$ prescribes the betting fraction in the $k$-th round using all available information: $\lambda_k=\Lambda_k(x;j_n,\dots,j_{k+1})$.
By composing individual steps like~\eqref{x_update}, we get the functional dependence $x_{k-1} =X_{k-1}(x;\Lambda_n,\dots,\Lambda_k;j_n,\dots,j_{k})$. Thus, the final amount may be written as $x_0=X_0(x,\Lambda,J)$, where $J=(j_n,\dots,j_1)$ is the whole sequence of chance events.

This game (or more exactly, the set of functions $(x,J)\mapsto X_0(x,\Lambda,J)$ corresponding to different strategies) is convex. Indeed, an interpolation between given strategies $\Lambda^{(0)}$, $\Lambda^{(1)}$ can be constructed by successively defining $\lambda^{(t)}_k =\Lambda^{(t)}_k(x;j_n,\dots,j_{k+1})$ as follows:
\begin{equation}
\lambda^{(t)}_k 
= \frac{(1-t) x^{(0)}_k \lambda^{(0)}_k + t x^{(1)}_k \lambda^{(1)}_k}
{x^{(t)}_k},
\end{equation}
where $x^{(t)}_n=x$, and $x^{(t)}_k$ for $k=n-1,\dots,0$ are obtained using the recurrence relation~\eqref{x_update}. Thus,
\begin{equation}
x^{(t)}_k = (1-t)x^{(0)}_k + tx^{(1)}_k\qquad
\text{for all $x$, $J$, and $k$}.
\end{equation}
The $k=0$ case of the last equation is exactly the desired convexity property.

We actually consider the $n$-round game in a bigger setting. In addition to the invested capital, the player has some side money. Without loss of generality, we take this amount to be $1$. No additional investments are allowed during the game. Thus, the total attained wealth is $w=1+x_0$. We are interested in maximizing its expected Bernoulli utility, $\EE[\ln(1+x_0)]$. More exactly, the goal is to calculate
\begin{equation}
f_n(x)=\max_{\Lambda}\EE_{J}\bigl[\ln(1+X_0(x,\Lambda,J))\bigr]
\end{equation}
and to find the corresponding optimal strategy. The latter depends only on the current amount rather than the full history, i.e.\ $\lambda_k=\Lambda_k(x_k)$.
 
A general solution is obtained using dynamic programming. Obviously, \begin{equation}
f_0(x)=\ln(1+x).
\end{equation}
For $k>0$, the situation in the $k$-th round is equivalent to a one-time gamble with utility function $f_{k-1}$. Hence,
\begin{equation}\label{iter_f}
\wideboxed{
f_{k}(x) = \max_{\lambda\in[0,1]} \sum_{j}p_j
f_{k-1}\bigl((1+\lambda(a_j-1))\,x\bigr),
}
\end{equation}
whereas $\Lambda_k(x)$ is given by the value of $\lambda$ at which the maximum is attained.

Equation~\eqref{iter_f} is easy to solve in asymptotic cases. For $x\to 0$, we may use the approximation $f_0(x)\approx x$. To maximize $\EE[x]$, one should bet all available money (because we have assumed that the gamble is favorable). We can now see, using induction in $k$, that $f_k$ is a linear function and that betting all the money is always good. (This is an example of a myopic strategy.) Indeed, if $f_{k-1}$ is linear, then the maximum in eq.~\eqref{iter_f} is attained at $\lambda=1$, and hence, $f_k(x)=\sum_j p_jf_{k-1}(a_jx)$ is also linear. More specifically, the solution is:
\begin{equation}
f_k(x)\approx \abar^kx\quad \text{for }x\to 0,\qquad
\text{where}\quad \abar=\sum_{j}p_ja_j>1.
\end{equation} 

In the $x\to\infty$ case, the utility function $f_0(x)$ may be approximated by $\ln x$ so that the optimal betting fraction is $\lambda_\Kelly=\lambda_*(0)$ (see section~\ref{sec:isoelastic}, particularly, figure~\ref{fig:opt_frac}). As in the previous case, the same myopic strategy can be used throughout the game. Thus, the expected utility grows at a constant rate $v_0$:
\begin{equation}\label{asymp_Kelly}
f_k(x)\approx \ln x + kv_0\quad \text{for }x\to \infty,\qquad
\text{where}\quad v_0 = \sum_{j}p_j\ln(1+\lambda_\Kelly(a_j-1)).
\end{equation} 

\begin{figure}
\centerline{\begin{tabular}{c@{\qquad}c}
\includegraphics{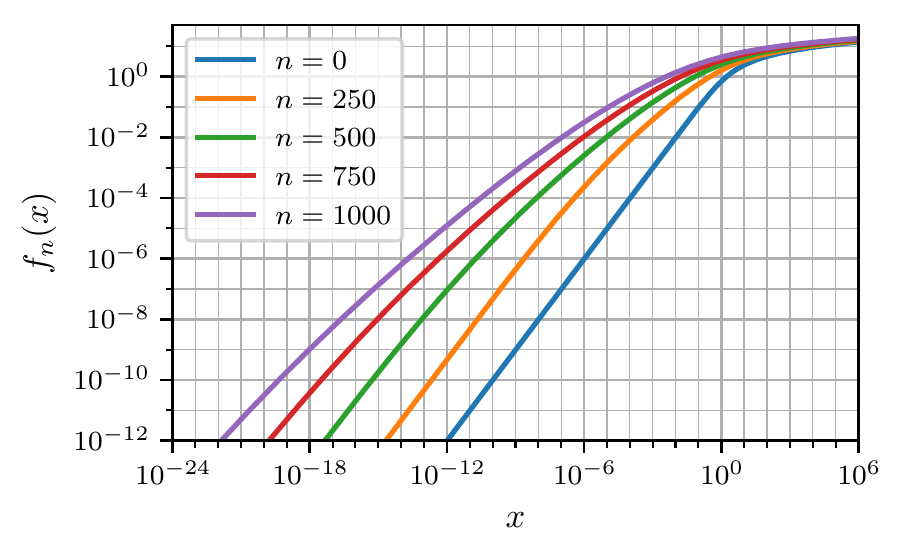} & \includegraphics{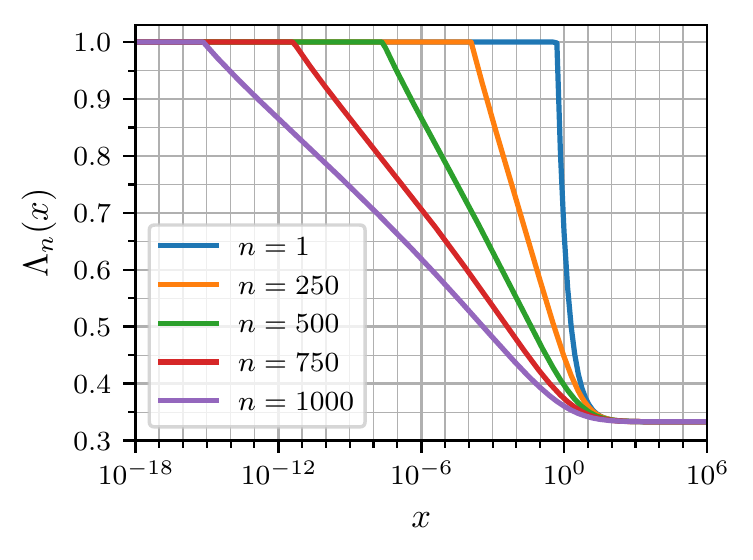}
\end{tabular}}
\caption{Numerical solution of equation~\eqref{iter_f} with the initial condition $f_0(x)=\ln(1+x)$. The second plot shows the optimal betting fractions, which vary from $1$ to $1/3$.}
\label{fig:results}
\end{figure}

\begin{figure}
\centerline{\begin{tabular}{c@{\qquad}c}
\includegraphics{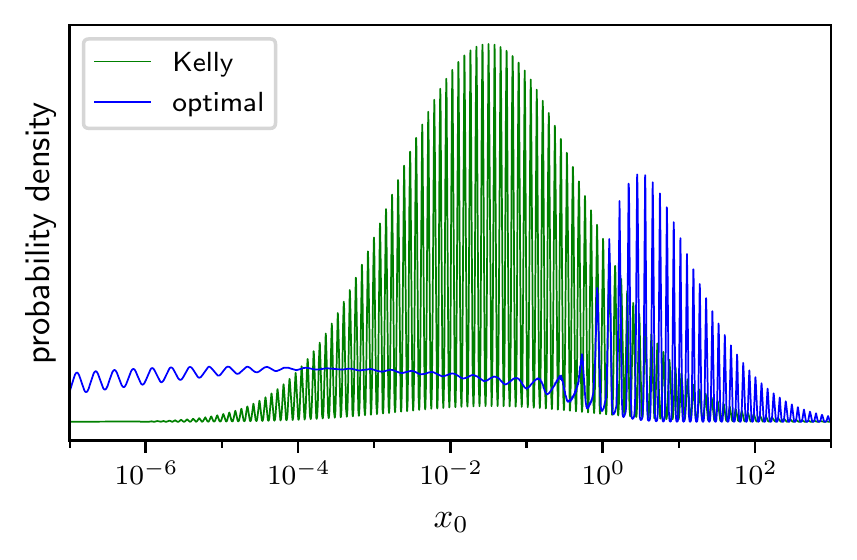} & \includegraphics{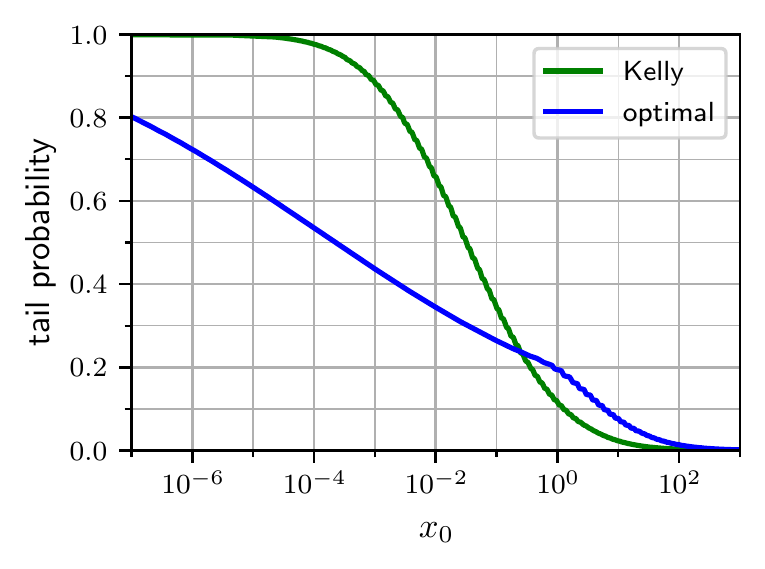}
\end{tabular}}
\caption{Probability distributions of the final capital $x_0$ when using the optimal strategy vs.\ Kelly's. The game consists of $n=1000$ rounds, and the initial capital is $x_n=10^{-3.3}$ (this choice of initial capital is discussed in the conclusion). The second plot shows the tail distribution, i.e.\ the probability that the final capital exceeds a given amount.}


\label{fig:probabilities}
\end{figure}

A numerical solution of equation~\eqref{iter_f} for our exemplary gamble (with $a_1=1.3$,\, $a_2=0.75$, and $p_1=p_2=0.5$) is shown in figure~\ref{fig:results}. Note that the plots cover a broad range of parameters, including some ridiculously small values of $x$. Furthermore, the expected utility $f_n(x)$ can be so small that it would not be worth the optimization effort. (In general, a tiny profit is worth pursuing only if it can be combined with other tiny profits.) However, the study of the problem for such extreme parameters helps one understand the large $n$ behavior, which is the subject of the next section. The example shown in figure~\ref{fig:probabilities} is less extreme. It illustrates the difference between the optimal and Kelly strategies for sufficiently large $n$ but intermediate $x$. The Kelly strategy guarantees a higher median value of the final capital $x_0$. The optimal strategy, on the other hand, achieves a higher probability of gaining an amount $x_0$ of the order of $1$. This probability can be small (in fact, it falls exponentially with $n$), but in this case, it is better to take the risk than having an assured but tiny gain. This will be explained in terms of an effective utility function with a suitably chosen risk parameter $0\le \alpha\le 1$.

We conclude this section with a rather technical remark. To solve equation~\eqref{iter_f} in a finite interval while avoiding an unreasonable computational cost, one has to impose some boundary conditions, for example, $f_n(0)=0$ and $f_n(x)=\ln x + nv_0$ for $x>x_{\mathrm{max}}$. A naive implementation of the second condition may cause an instability. To avoid this and other possible instabilities, we maintained an invariant of the exact problem---that the function $f_n$ is concave for all $n$. More exactly, if $f_{n-1}$ is concave, then $f_n$ is concave; this property is closely related to the convexity of the game itself. To reconstruct the function from its grid values, we used a concavity-preserving quadratic interpolation. When gluing the numerical solution for $x<x_{\mathrm{max}}$ and the Kelly asymptotics for $x>x_{\mathrm{max}}$, we cut the ``tooth'' at the boundary so that the resulting function is concave.

\section{Asymptotic regions and the WKB solution}

\begin{figure}
\centerline{\begin{tabular}{c@{\hspace{1.5cm}}c}
\(
\vcenter{\hbox{\includegraphics{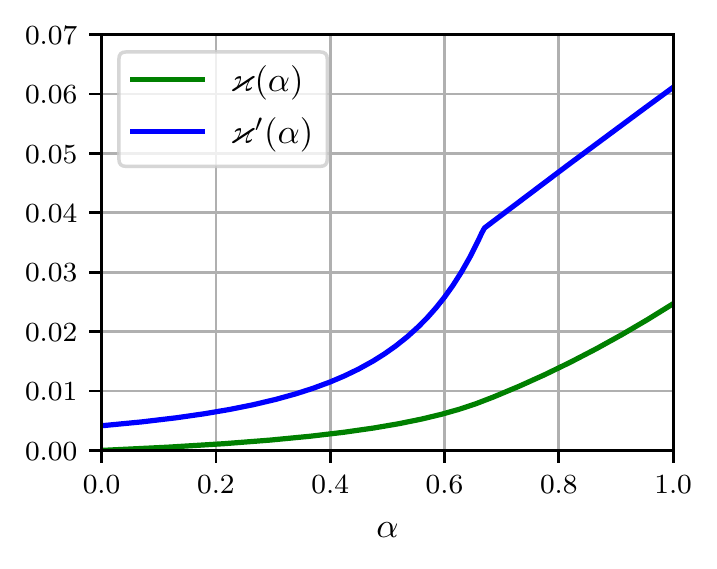}}}
\) &
\(\def\arraystretch{1.3}
\begin{array}{|c|r@{}l|}
\hline
\text{time} & t&{}=in\\
\text{spatial coordinate} & q&{}=\ln x\\
\text{momentum} & p&{}=-i\alpha\\
\text{Hamiltonian} & H&{}=\kap(\alpha)\\
\text{velocity} & \frac{dq}{dt}=\frac{\partial H}{\partial p}
&{}=i\kap'(\alpha)\\[3pt]
\hline
\end{array}
\)\\[5pt]
a) & b)
\end{tabular}}
\caption{a)~Growth rate $\kap$ and its derivative $\kap'$ as functions of the risk parameter $\alpha$. b)~Mapping between quantum mechanics of a 1d particle and the gambling problem.}
\label{fig:kappa}
\end{figure}

Let us now find the asymptotic form of the functions $f_n$ at large $n$. The key insight, which can be gleaned from the numerics, is that $\frac{d\ln f_n(x)}{d\ln x}$ (i.e.\ the slope of the curves in the left plot in figure~\ref{fig:results}) varies slowly over a broad range of $x$. For the purpose of optimizing the betting fraction at a given point $x=x_n$, we may assume the slope $\alpha$ to be constant. That is, we may use the approximation
\begin{equation}\label{power_approx}
f_{n-1}(x)\approx c_{n-1}x^{\alpha},\qquad 0\le \alpha\le 1
\end{equation}
in some neighborhood of the point $x_n$. Power-law functions are invariant under the iteration by equation~\eqref{iter_f}, with the overall factor growing exponentially. More exactly, $c_n=e^{\kap(\alpha)}c_{n-1}$, where
\begin{equation}\label{kap_r}
\kap(\alpha) = \max_{\lambda\in[0,1]} r(\alpha,\lambda),\qquad
r(\alpha,\lambda)=\ln\biggr(\sum_{j}p_j(1+\lambda(a_j-1))^{\alpha}\biggl).
\end{equation}
The function $\kap(\alpha)$ and its derivative $\kap'(\alpha)$ for the specific gamble are plotted in figure~\ref{fig:kappa}a. We will show that $\kap$ is convex, and hence, $\kap'$ is monotone.

When using the approximation~\eqref{power_approx} locally, the equation $c_n=e^{\kap(\alpha)}c_{n-1}$ should be replaced with this one:
\begin{equation}\label{WKB}
\frac{\partial \ln f_n(x)}{\partial n}
= \kap(\alpha_n(x))\qquad
\text{with}\quad \alpha_n(x) = \frac{\partial\ln f_n(x)}{\partial\ln x}.
\end{equation}
Here $n$ is treated as a continuous variable, the second variable being $q=\ln x$. This is, essentially, the WKB approximation used in a slightly unusual situation. Let us digress a bit and elaborate on this analogy. 

The WKB approximation is commonly used in quantum mechanics. It amounts to writing the wavefunction of a particle moving along the $q$ axis as $\psi(t,q) =e^{iS(t,q)}$, up to some factor that varies slowly in space and time. Then the Schr\"odinger equation for $\psi$ is reduced to the Hamilton-Jacobi equation, $\frac{\partial S}{dt} =-H\bigl(q,p)$ with $p=\frac{\partial S}{\partial q}$. One can readily see the analogy with equation~\eqref{WKB}; the variable mapping between the two problems is shown in figure~\ref{fig:kappa}b. Note that the momentum $p=-i\alpha$ is imaginary, like in the quantum tunneling problem. Studying quantum evolution in imaginary time is also a common trick, used to calculate thermodynamic properties.

Returning to the main subject, equation~\eqref{WKB} is first-order, and thus, can be solved by the method of characteristics. Characteristics are lines of constant $\alpha$ in the $(\ln x, n)$ plane. They are given by the equation
\begin{equation}\label{char_eq}
\frac{d\ln x}{dn} = -\kap'(\alpha).
\end{equation}
Indeed, let us differentiate both sides of the first equation in~\eqref{WKB} with respect to $\ln x$ and express $\frac{\partial \ln f_n(x)}{\partial\ln x}$ on the left-hand side as $\alpha=\alpha_n(x)$. The result is $\frac{\partial\alpha}{\partial n} =\kap'(\alpha)\frac{\partial\alpha}{\partial\ln x}$, implying that $\alpha$ is constant on lines~\eqref{char_eq}. These lines are projected from points $x_0$ on the $n=0$ line, where $\alpha=\alpha_0(x_0)$ is as follows:
\begin{equation}
\alpha_0(x) = \frac{d\ln f_0(x)}{d\ln x}= \frac{x}{(1+x)\ln(1+x)} \to
\begin{cases}
1 &\text{if } x\to 0,\\
0 &\text{if } x\to \infty.
\end{cases}
\end{equation}
On each characteristic, equation~\eqref{WKB} is reduced to an ordinary differential equation, which has a simple solution:
\begin{equation}\label{WKB_sol}
f^{\mathrm{WKB}}_n\bigl(e^{-n\kap'(\alpha)}x_0\bigr)
= e^{-n(\alpha\kap'(\alpha)-\kap(\alpha))}f_0(x_0),\qquad
\text{where}\quad \alpha=\alpha_0(x_0).
\end{equation}

\begin{figure}
\centerline{\begin{tabular}{c@{\hspace{1.5cm}}c}
\includegraphics{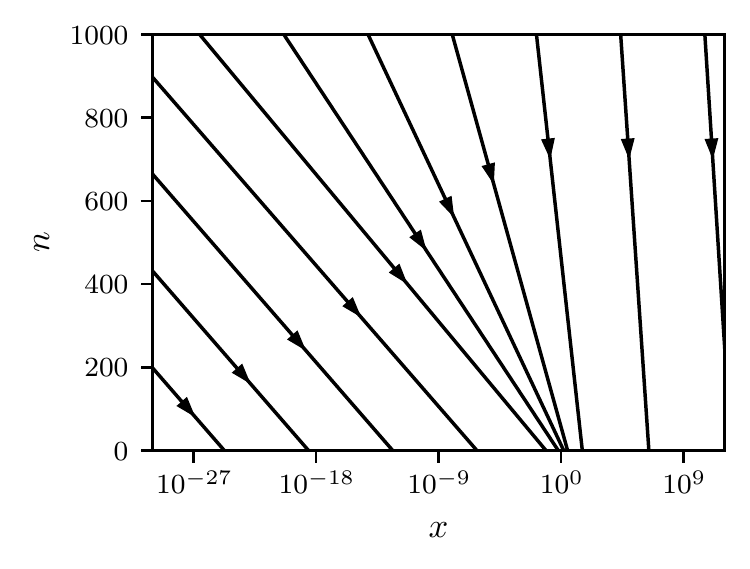} & \raisebox{15pt}{\includegraphics{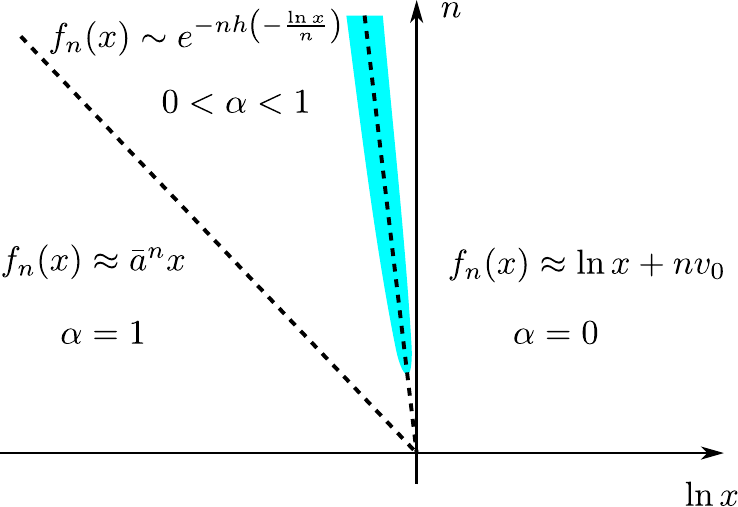}}
\end{tabular}}
\caption{WKB characteristics (left) and asymptotic regions in the $n\to\infty$ limit (right). The blue area in the left figure will be studied in section~\ref{sec:diffusion}.}
\label{fig:regions}
\end{figure}

From the algorithmic point of view, $f^{\mathrm{WKB}}_n(x)$ is computed by shooting a characteristic with a suitable slope $-\kap'(\alpha)$ from the point $(\ln x, n)$  so as to satisfy the equation $\alpha=\alpha_0(x_0)$ at the endpoint, see figure~\ref{fig:regions}. This is an explicit expression for $\kap'(\alpha)$:
\begin{equation}
\kap'(\alpha)
= \frac{\partial r(\alpha,\lambda)}{\partial\alpha}
\biggr|_{\lambda=\lambda_*(\alpha)}
= \frac{\sum_{j}p_j\tilde{a}_j^{\alpha}\ln\tilde{a}_j}
{\sum_{j}p_j\tilde{a}_j^{\alpha}},\qquad
\text{where}\quad \tilde{a}_j=1+\lambda_*(\alpha)\,(a_j-1).
\end{equation}
In particular, $\kap'(0)$ is equal to the Kelly rate $v_0$ defined in equation~\eqref{asymp_Kelly}, and we will denote $\kap'(1)$ by $v_1$. Let us use the following approximation:
\begin{equation}
\alpha_0(x) \approx
\begin{cases}
1 &\text{if } x<0,\\
0 &\text{if } x>0,\\
\text{any number between $0$ and $1$} &\text{if } x=0.
\end{cases}
\end{equation}
It incurs error in a relatively small interval, $x\sim 1$, and the overall precision loss is comparable to that due to the WKB approximation itself. Thus, we arrive at the asymptotic region picture shown in figure~\ref{fig:regions}. In the intermediate region, located between the lines with slopes $-v_1$ and $-v_0$, the approximate solution is
\begin{equation}
f_n(x)\sim e^{-nh\left(-\frac{\ln x}{n}\right)},
\end{equation}
where $h$ the Legendre transform of the function $\kap$:
\begin{equation}\label{h_def}
h(v)=\max_{\alpha\in[0,1]}(\alpha v-\kap(\alpha))\qquad
\text{for}\quad v_0\le v\le v_1.
\end{equation}
That is, $h(v)=\alpha\kap'(\alpha)-\kap(\alpha)$ for the unique $\alpha\in[0,1]$ satisfying the equation $\kap'(\alpha)=v$.

Let us now establish some useful properties of the functions $\kap(\alpha)$,\, $r(\alpha,\lambda)$ (see equation~\eqref{kap_r}), and $h(v)$. Some of the subsequent arguments are borrowed from statistical mechanics and the derivation of Chernoff's bound.

\begin{Proposition}
The functions $\kap(\alpha)$ and $r(\alpha,\lambda)$ are convex in $\alpha$.
\end{Proposition}
\begin{proof}
Let $\Delta$ be the set of probability distributions on the outcomes of a single gamble. Then
\begin{equation}\label{r_as_max}
r(\alpha,\lambda) = \max_{q\in\Delta}
\biggl(\sum_j q_j\ln\frac{p_j\tilde{a}_j^\alpha}{q_j}\biggr),\qquad
\text{where}\quad \tilde{a}_j=1+\lambda(a_j-1).
\end{equation}
Indeed, the maximum is attained at an interior point of $\Delta$, which can be determined by taking partial derivatives with respect to $q_j$ and using a Lagrange multiplier. Specifically, $q_j=p_j\tilde{a}_j^\alpha/A$ with $A=\sum_{j}p_j\tilde{a}_j^\alpha$, and hence, $\sum_j q_j \ln\frac{p_j\tilde{a}_j^\alpha}{q_j} =\ln A=r(\alpha,\lambda)$.

Next, we interpret the maximum in equation~\eqref{r_as_max} as a Legendre transform of some function in $\alpha$:
\begin{equation}
r(\alpha,\lambda)=\max_{v}(\alpha v - s(v,\lambda)),
\end{equation}
where
\begin{equation}\label{rel_entropy}
s(v,\lambda)=\min_{q\in Q(v)}
\biggl(\sum_{j}q_j\ln\frac{q_j}{p_j}\biggr),\qquad
Q(v)=\biggl\{q\in\Delta:\,\sum_{j}q_j\ln\tilde{a}_j=v\biggr\}.
\end{equation}
(The minimum over an empty set is defined as $+\infty$.) It follows that $r(\alpha,\lambda)$ is convex in $\alpha$, and hence, $\kap(\alpha)=\max_{\lambda\in[0,1]}r(\alpha,\lambda)$ is also convex.
\end{proof}

\begin{Proposition}
The function $h$ defined by equation~\eqref{h_def} admits the following representation:
\begin{equation}
h(v)=\min_{\lambda\in[0,1]}s(v,\lambda)\qquad
\text{for}\quad v_0\le v\le v_1,
\end{equation}
where $s(v,\lambda)$ is the minimum relative entropy, see equation~\eqref{rel_entropy}.
\end{Proposition}

\begin{proof}
It follows from the convexity of the function $q \mapsto\sum_jq_j\ln\frac{q_j}{p_j}$ that $s(v,\lambda)$ is convex in $v$. Applying the Legendre transform to a convex function twice gives the same function; therefore,
\begin{equation}\label{s_v_lambda}
s(v,\lambda)=\sup_{\alpha\in\RR}(\alpha v -r(\alpha,\lambda)).
\end{equation}

Let us temporarily restrict $\alpha$ to the interval $[0,1]$. If $\alpha\in[0,1]$, then $r(\alpha,\lambda)$ is concave in $\lambda$, and thus, we can apply von Neumann's minimax theorem~\cite{vNmo44}:
\begin{equation}
h(v)= \max_{\alpha\in[0,1]}\min_{\lambda\in[0,1]}(\alpha v-r(\alpha,\lambda))
=\min_{\lambda\in[0,1]}\max_{\alpha\in[0,1]}(\alpha v-r(\alpha,\lambda)).
\end{equation}
We claim that the constraint $\alpha\in[0,1]$ on the last maximum can be dropped, provided $v_0\le v\le v_1$. To see this, let us assume that both inequalities are strict; the general case follows by continuity. If $v_0<v<v_1$, the maximin is attained at some point $(\alpha(v),\lambda(v))$ with $0<\alpha(v)<1$. This point is a saddle of the function $\ph(\alpha,\lambda)=\alpha v-r(\alpha,\lambda)$, meaning that
\begin{equation}
\min_{\lambda\in[0,1]}\ph(\alpha(v),\lambda) = \ph(\alpha(v),\lambda(v))
= \max_{\alpha\in[0,1]}\ph(\alpha,\lambda(v)).
\end{equation}
Since $0<\alpha(v)<1$ and $\ph(\alpha,\lambda(v))$ is concave in $\alpha$ for $\alpha\in\RR$, the last maximum is, actually, global. Thus, $(\alpha(v),\lambda(v))$ satisfies the saddle condition for $\lambda\in[0,1]$,\, $\alpha\in\RR$, and hence, is also a minimax point in this domain. We conclude that $h(v) =\min_{\lambda\in[0,1]} \sup_{\alpha\in\RR}(\alpha v-r(\alpha,\lambda))$, where the supremum is equal to $s(v,\lambda)$ due to equation~\eqref{s_v_lambda}.
\end{proof}

WKB characteristics have an interesting probabilistic interpretation. Its exact statement would have to be elaborate and the proof would involve a Chernoff bound, with the relative entropy playing its usual role. We will not attempt that but notice that this qualitative picture explains the numerics well. Let $x_n$ be the initial capital, and let $p(x_0)$ denote the tail distribution of the final capital, assuming that the player uses the optimal strategy. Then the expected utility $f_n(x_n)=\EE[f(x_0)]$ can be expressed as follows:
\begin{equation}
f_n(x_n)=\int f_0(x_0)\left(-\frac{dp(x_0)}{dx_0}\right)dx_0.
\end{equation}
We claim that this integral is dominated by a small neighborhood of the endpoint of the characteristic passing through $(\ln x_n,n)$. Given that, the characteristic represents an ``optimistic trajectory'' the player hopes to follow. The expected utility along this trajectory increases exponentially at the rate $-\frac{d\ln f_n(x)}{dn}=\alpha\kap'(\alpha)-\kap(\alpha)$. Here, we refer to the conditional expectation value, assuming that the player stays on track. However, the probability of this event decreases at the same rate, such that the total expectation value is conserved. (The probability could drop even faster if there were other paths to the destination. However, if the player ever falls behind, the chances to recover are very slim.) Thus, we may call $v=\kap'(\alpha)$ the ``optimistic growth rate'' and $h(v)=\alpha\kap'(\alpha)-\kap(\alpha)$ the ``failure rate''. Our asymptotic analysis is all about the balance between these quantities, as is evident from this expression for the WKB solution~\eqref{WKB_sol}:
\begin{equation}
\wideboxed{
f^{\mathrm{WKB}}_n(x)=\max_{v}e^{-nh(v)}f_0\bigl(e^{nv}x\bigr).
}
\end{equation}

\section{Diffusion approximation}\label{sec:diffusion}

\begin{figure}
\centerline{\begin{tabular}{c@{\hspace{1.5cm}}c}
\includegraphics{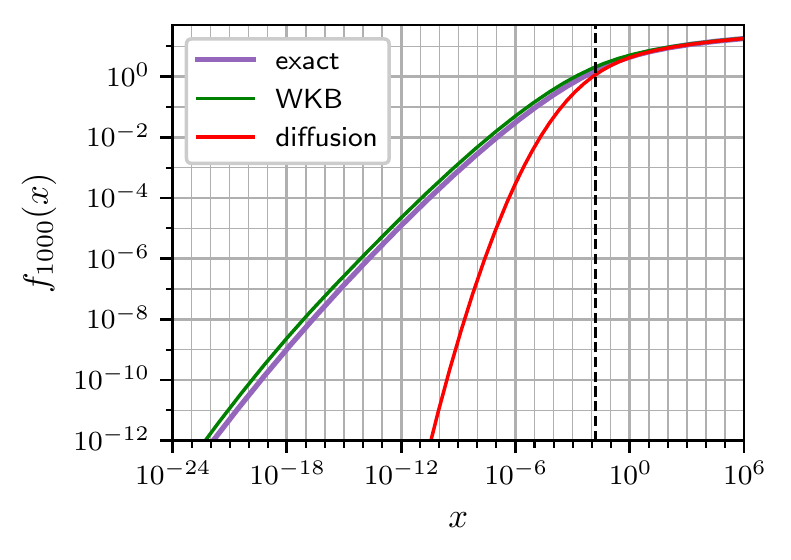} & \includegraphics{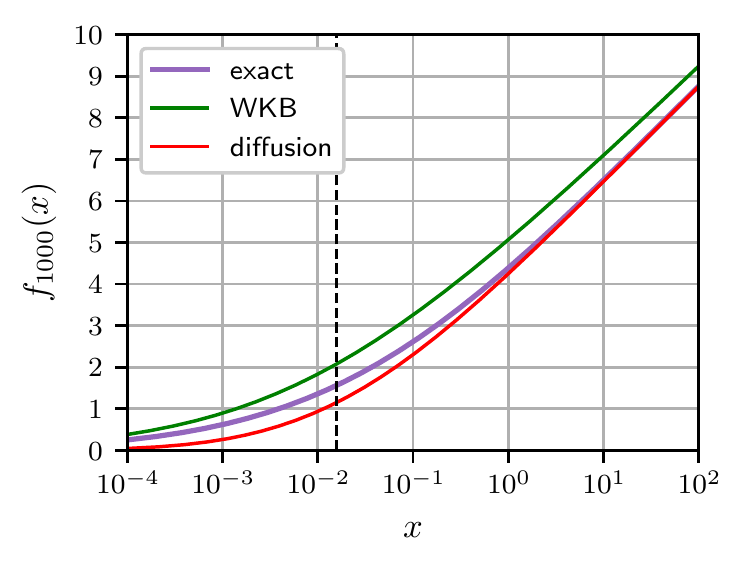}
\end{tabular}}
\caption{Comparison of the exact solution $f_n$ with those obtained using the WKB and diffusion approximations for $n=1000$. The dashed vertical line is positioned at $e^{-nv_0}$, which corresponds to the boundary of the $\alpha=0$ region.}
\label{fig:comparison}
\end{figure}

As already mentioned, the expected gains $f_n(x)$ in the $\alpha=1$ region and the intermediate $0<\alpha<1$ region decrease exponentially with $n$. Let us now study the boundary between the intermediate region and the Kelly region (with $\alpha=0$), where the gains can be significant. The Kelly region will also be covered by our analysis. The plan is to expand $\kap(\alpha)$ to the second order in $\alpha$ and write the corresponding differential equation, which is, essentially, the diffusion equation.

We begin with a quadratic expansion of $r(\alpha,\lambda)$ in $\alpha$:
\begin{equation}
r(\alpha,\lambda) = \alpha\,\bigl(
v_0(\lambda)+ D(\lambda)\,\alpha + O(\alpha^2)\bigr),
\end{equation}
where
\begin{equation}
v_0(\lambda)=\sum_{j}p_j\ln\tilde{a}_j,\qquad
D(\lambda)= \frac{1}{2} \biggl(\sum_{j}p_j(\ln\tilde{a}_j)^2
-\Bigl(\sum_{j}p_j\ln\tilde{a}_j\Bigr)^2\biggr),\qquad
\tilde{a}_j=1+\lambda(a_j-1).
\end{equation}
The following intermediate calculations use the assumption $\lambda_\Kelly=\lambda_*(0)<1$, but the result is also valid for $\lambda_\Kelly=1$. Expanding $v_0(\lambda)$ and $D(\lambda)$ in $\lambda-\lambda_\Kelly$, we go one order higher than necessary:
\begin{equation}
\begin{aligned}
v_0(\lambda)&=v_0(\lambda_\Kelly)
+v_0'(\lambda_\Kelly)\,(\lambda-\lambda_\Kelly)
+\frac{v''(\lambda_\Kelly)}{2}\,(\lambda-\lambda_\Kelly)^2 +\cdots,\\[2pt]
D(\lambda)&=D(\lambda_\Kelly)
+D'(\lambda_\Kelly)\,(\lambda-\lambda_\Kelly) +\cdots.
\end{aligned}
\end{equation}
The higher accuracy helps to illustrate how the $\lambda$ optimization works. Note that $v_0'(\lambda_\Kelly)=0$ because by definition, $\lambda_\Kelly$ is the point at which $v_0(\lambda) =\sum_{j}p_j\ln(1+\lambda(a_j-1))$ attains a maximum value. Thus,
\begin{equation}\label{optfrac_exp}
\lambda_*(\alpha)=\lambda_\Kelly
-\frac{D'(\lambda_\Kelly)}{v''(\lambda_\Kelly)}\,\alpha
+O(\alpha^2).
\end{equation}
However, the approximation $\lambda_*(\alpha)\approx\lambda_\Kelly$ will be sufficient for our purposes, and it is also applicable if $\lambda_\Kelly=1$. Using either this approximation or equation~\eqref{optfrac_exp}, we get:
\begin{equation}\label{kap_alpha_2}
\kap(\alpha)=v_0\alpha+D\alpha^2+O(\alpha^3),\qquad
\text{where}\quad v_0=v_0(\lambda_\Kelly),\quad D=D(\lambda_\Kelly).
\end{equation}

Independently of the concrete expression for $\kap(\alpha)$, the approximation $\lambda\approx\lambda_\Kelly$ makes the problem linear:
\begin{equation}\label{recrel_approx}
f_{n}(x) \approx \sum_{j}p_jf_{n-1}(\tilde{a}_jx),\qquad
\tilde{a}_j=1+\lambda_\Kelly(a_j-1).
\end{equation}
It is convenient to write it in lin-log coordinates, i.e.\ in terms of the variable $y=\ln x$ and the function
\begin{equation}
g_n(y)=f_n(e^y).
\end{equation}
Thus, the recurrence relation~\eqref{recrel_approx} becomes
\begin{equation}
g_n(y) \approx \sum_{j}p_j\,g_{n-1}(y+\ln\tilde{a}_j).
\end{equation}
To simplify it further, we will fit the set of standard solutions $g_{n}(y)=e^{\kap(\alpha)n+\alpha y}$ with a linear second-order differential equation using the expression~\eqref{kap_alpha_2} for $\kap(\alpha)$:
\begin{equation}
\frac{\partial g}{\partial n}
= v_0\,\frac{\partial g}{\partial y} + D\,\frac{\partial^2 g}{\partial y^2}.
\end{equation}
Its solution can be written explicitly:
\begin{equation}
g_n(y)=\int K_n(y,z)\,g_0(z)\,dz,\qquad
\text{where}\quad
K_n(y,z)=\frac{1}{2\sqrt{\pi Dn}}\,\exp\biggl(-\frac{(y-z+v_0n)^2}{4Dn}\biggr).
\end{equation}
Replacing $g_0(y)=\ln(1+e^y)$ with a crude approximation,
\begin{equation}
g_0(y)=y\,\theta(y),
\end{equation}
we get:
\begin{equation}\label{diff_sol}
g_n(y)=\sqrt{4Dn}\: g\biggl(\frac{y+v_0n}{\sqrt{4Dn}}\biggr),\qquad
g(t) = t\,\frac{1+\erf(t)}{2} + \frac{e^{-t^2}}{2\sqrt{\pi}},
\end{equation}
where $\theta$ is the step function and $\erf$ is the error function.

\section{Conclusion}

We are now equipped with the tools to return to the original question and determine how one should realistically play this game, specifically considering our youngest likely readers. We first estimate the external capital for this reader, who we place as a grad student. There are approaches we can take to estimate this. The EPA puts the value of a statistical life at $\$7.4$ million in 2006 dollars~\cite{EPA-VSL}, which adjusts to $\$9.5$ million. However, this estimate factors in death and pain, rather than just monetary losses.

Better suited to our purposes is a result from a paper by Carnevale, Rose, and Cheah~\cite{CRC11} which places the median lifetime earnings of Americans with a master's degree at $\$2.7$ million in 2009 dollars, equivalent to roughly $\$3.2$ million now. However, this is still likely an overestimate for two main reasons: first, working has direct costs (gas costs of driving to work, housing costs of living near job centers, etc.), and second, doing no job for some amount of money is worth more than doing a job for that amount of money (in addition to the risks involved in potentially losing a job).

Instead, we can pose a thought experiment: how much money would one take in exchange for giving up all future earnings (except perhaps passive investment)? We can think about this using annuities since they are exactly the spreading out of a lump sum over a lifetime. A Charles Schwab estimation tool~\cite{CS-calculator} gives an annuity income of roughly $\$5{,}000$ per month for a 25-year-old with $\$2{,}000{,}000$ initial investment. This is substantially less than the expected income, but we must consider that this already factors in retirement savings (since the annuity makes no distinction), which brings it up to the equivalent of roughly $\$70{,}000$ annually (assuming a fairly typical $15\%$ retirement savings rate). Considering this is income one gets never working again (i.e. an extremely early retirement), and that this game is best viewed on a logarithmic scale (so the difference between $\$2$ and $\$3$ million is minor), we will use $\$2{,}000{,}000$ as our estimate for external capital.

Since rather than solving $\ln(x_{ext} + x_{int})$, we are normalizing to $\ln(1 + x)$, our initial $x_n$ is $\frac{\$1{,}000}{\$2{,}000{,}000}=5 \cdot 10^{-4} \approx 10^{-3.3}$,\, $n=1{,}000$. Solving \eqref{iter_f} with dynamic programming, we see that the optimal betting fraction is roughly $0.494$ for the first move. If that coin-flip is good, our capital increases and our next bet drops slightly to $0.492$ of our total capital. If it's bad, our bet increases similarly. In general, the optimal bet will increase to infinity (but is capped at $1$ by the rules) as you have less to lose while decreasing towards $\frac{1}{3}$ (the $\ln(x)$ optimization) as your internal capital overwhelms external capital. As a consequence, bad luck in this game reinforces itself but good luck will encourage you to be more cautious, increasing the chances for further success.

\begin{figure}
\centerline{\begin{tabular}{c@{\qquad}c}
\includegraphics{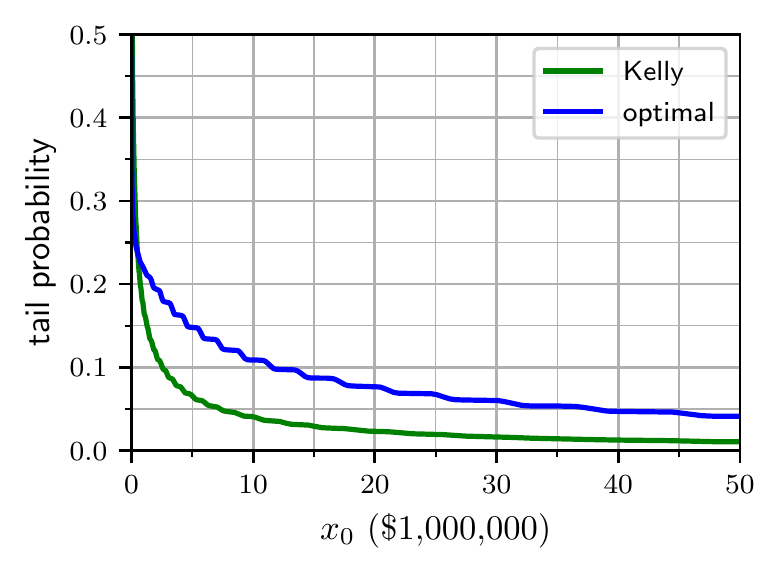}
\end{tabular}}
\caption{Linear scale tail distribution of the final capital $x_0$ (in millions of dollars). The game consists of $n=1000$ rounds and the external capital is $\$2{,}000{,}000$.}

\label{fig:probabilities_linear}
\end{figure}

We can summarize from figure~\ref{fig:probabilities} that optimizing $\ln(1 + x)$ leads to ``double or nothing'' behavior: it's a waste of effort to go for a mediocre result and instead optimal to risk the small amount in play for results on order with external capital. We can see in figure~\ref{fig:probabilities_linear} that the probabilities of large sums are substantially higher under optimal play.

These results stand in stark contrast to the conventional wisdom that super-Kelly betting is a path to ruin. We see that, in a situation where there is capital that cannot be applied, betting more aggressively than the Kelly strategy is a sound strategy. Because everyone has capital they cannot apply, whether it be explicit non-liquid investments or simply future productivity, we conclude that super-Kelly betting can be a reasonable option.

\section*{Acknowledgments}

We thank Gregory Falkovich for useful comments. A.K.\ is supported by the Simons Foundation under grant~376205 and by the Institute of Quantum Information and Matter, a NSF Frontier center funded in part by the Gordon and Betty Moore Foundation.

\bibliography{Kelly-opt}
\bibliographystyle{JHEP-mod}

\end{document}